\documentclass[
aps,reprint,amsmath,amssymb]{revtex4-1}
\usepackage[utf8]{inputenc}
\usepackage{graphicx}
\usepackage[dvipsnames]{xcolor}
\usepackage[sort&compress]{natbib}

\begin{document}

\preprint{APS/123-QED}

\title{An Analytical Solution to the $k$-core Pruning Process}

\author{Gui-Yuan Shi$^{1,*}$}
\author{Rui-Jie Wu$^{1,*,+}$}
\author{Yi-Xiu Kong$^{1,*}$}
\author{H. Eugene Stanley$^{2,+}$}
\author{Yi-Cheng Zhang$^{1}$}
\affiliation{$^{1}$Department of Physics, University of Fribourg, Fribourg 1700, Switzerland.}
\affiliation{$^{2}$Center for Polymer Studies and Department of Physics, Boston University, Boston, MA 02215, USA.} 
\affiliation{$^{*}$These authors contributed equally to this work.}
\affiliation{$^{+}$To whom correspondence should be addressed:
Email: ruijie.wu@unifr.ch (R.-J.W.), hes@bu.edu (H.E.S.) }

\date{\today}

\begin{abstract}
$k$-core decomposition is widely used to identify the center of a large network, it is a pruning process in which the nodes with degrees less than $k$ are recursively removed. Although the simplicity and effectiveness of this method facilitate its implementation on broad applications across many scientific fields, it produces few analytical results. We here simplify the existing theoretical framework to a simple iterative relationship and obtain the exact analytical solutions of the $k$-core pruning process on large uncorrelated networks. From these solutions we obtain such statistical properties as the degree distribution and the size of the remaining subgraph in each of the pruning steps. Our theoretical results resolve the long-lasting puzzle of the $k$-core pruning dynamics and provide an intuitive description of the dynamic process.
\end{abstract}
\maketitle

\section*{Introduction}

\begin{figure}[htb]
\centering
\includegraphics[width=8.6cm]{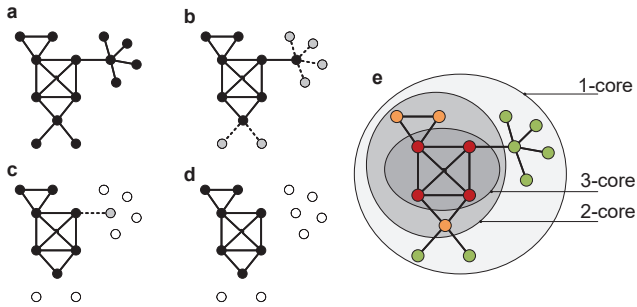}
\caption{
\textbf{The illustration of $k$-core pruning process and $k$-core.} \textbf{a-d}, the detailed pruning steps of $2$-core, the discs represent nodes and the solid lines indicate the links. Grey discs and dashed lines represent the nodes and links that will be removed in this step. Dashed circles stand for the nodes that have been removed. \textbf{a)}The starting network. \textbf{b)}The first pruning step, remove the nodes with degrees less than $2$, and links incident to them. Note after this step, the node in the top right corner that previously had degree of $5$ will be left with only one link. \textbf{c)}The second pruning step, remove the nodes whose degree is fewer than $2$. \textbf{d)}The final step. All nodes remained in the network have more than $2$ links, and the iterative process terminates in this step. \textbf{e)}$k$-core decomposition result of the network. $1$-core contains all the nodes, so it is trivial. $2$-core contains the nodes marked with red and orange, and $3$-core of the network contains the red nodes. 
}
\label{kcore}
\end{figure}

In the k-core pruning process we recursively
remove the nodes with degrees less than $k$. We iteratively repeat the
process until a finite-sized subgraph, the $k$-core of the network, is
obtained. If it is not obtained, the network
disappears. Fig.~\ref{kcore} shows a simplified picture of $k$-core
decomposition. Note that a $2$-core decomposition is performed on the
network, and that the final $2$-core is obtained in step 3 (Fig.~\ref{kcore}\textbf{d}).

The first proposed application of $k$-core decomposition was to measure
the centrality of nodes in a network~\cite{seidman1983network}, but more
recently it has been applied to many disciplines, including biology,
informatics, economy, and network science. Bader et
al.~\cite{bader2003automated} developed an algorithm based on $k$-core
decomposition to identify the densely linked regions in the
protein-protein interaction (PPI) network, which may represent molecular
complexes. Altaf et al. ~\cite{altaf2003prediction} use the k-core decomposition 
to predict the PPI functions of several function-unknown proteins. Stefan Wuchty et
al.~\cite{wuchty2005peeling} discovered that the probability of proteins
being essential and conserved through stages of evolution increases with
the $k$-coreness of the protein. Nir Lahav et al.~\cite{lahav2016k} used
$k$-core decomposition to describe the hierarchical structure of the
cortical organization in the human brain. They discovered that the
strongest hierarchy serves as a platform for the emergence of
consciousness. Researchers in information science, economics, and
complex networks have also used $k$-core as a filter to obtain relevant
information in a large system~\cite{gaertler2004dynamic,
  carmi2007model}, to identify the central countries during economic
crises~\cite{garas2010worldwide}, to locate the most influential
propagators in a complex network system~\cite{kitsak2010identification,morone2015influence},
and to predict structural collapse in mutualistic ecosystems~\cite{morone2019k}.

Because $k$-core decomposition has so many applications, researchers
\cite{fernholz2004cores,dorogovtsev2006k} have used theoretical analyses
and numerical simulations to study the final state of $k$-core
decomposition and have found that a giant $k$-core emerges in the form
of a phase transition \cite{dorogovtsev2008critical}.  The $k$-core
exists only when the initial average degree of nodes is above a critical
point, denoted $c^*$. Baxter et al.~\cite{baxter2015critical} propose a
theoretical framework of four equations to describe the evolution of the
degree distribution. Their numerical calculation on Erd\H{o}s-R\'enyi
networks when $k=3$ show that a long-lasting transient ``plateau'' stage
exists before the final collapse when the initial mean degree is close
to the critical value.

Although the numerical result reveals that the $k$-core pruning process
has many interesting properties, the analytical result is still hindered
by its mathematical difficulty \cite{baxter2015critical}. In this paper
we correct an oversight in the previous research
\cite{baxter2015critical} in which the probability cannot be normalized
because it is missing a non-negligible term in its original theoretical
framework. More important, we solve the mathematical problem by inducing
an auxiliary series and obtaining the analytical solution for any large
uncorrelated network. Our results clearly describe the subgraph in each
pruning step and agree with various numerical simulations. 

To analyze of $k$-core decomposition theoretically we need to determine how many nodes remain in the network after each pruning process and discover the structural topology, i.e., the degree distribution, of the subgraph. We here present complete results for both. We begin with a brief introduction of the theoretical framework given in Ref.~\cite{baxter2015critical}. 

In each step of the $k$-core pruning process on a large uncorrelated
network with a finite average degree, we remove nodes with a degree less
than $k$. For convenience we assume that pruned nodes remain in the
network, but are with degrees equal to zero. 
We denote the $n^{th}$ network after pruning $n$ by $\mathcal{N}_n$, 
and designate $G_{n,0}(z)=\sum_{j=0}^\infty p_{n,j}z^j$---the notation supplied in Ref.~\cite{newman2018networks}---to be the probability generating function for the degree distribution of $\mathcal{N}_n$,
where $p_{n,j}$ is the degree distribution in $\mathcal{N}_n$. 
Similarly $G_{n,1}(z)=\sum_{j=0}^\infty q_{n,j}z^j$ is the probability generating
function for the excess degree distribution of $\mathcal{N}_n$, where
$q_{n,j}$ is the excess degree distribution, 
i.e., the degree distribution of a node reached by following a randomly chosen link, excluding the chosen link itself.  
We simplify the notation of initial generating functions from $G_{0,0}$ and $G_{0,1}$ to $G_0$ and $G_1$, respectively.

In pruning process $n$ on network $\mathcal{N}_{n-1}$ 
(see SI 1A for a schematic illustration), $v_{n-1}$
is the probability that when we randomly follow a link to one node in
$\mathcal{N}_{n-1}$ it has a degree greater than $k-2$,
\begin{equation}
v_{n-1}=1-\sum_{j=0}^{k-2}q_{n-1,j}.
\end{equation}
Here the set of nodes with a degree equal to $0$ after $k$-core pruning
$n$ has (i) nodes with a degree less than $k$ and (ii) nodes with
degrees not less than $k$ but with neighbors all having degrees less
than $k$. Note that in previous research \cite{baxter2015critical} this
latter term was missing,
\begin{equation}
p_{n,0}=\sum_{j=0}^{k-1}p_{n-1,j}+\sum_{j=k}^{\infty}p_{n-1,j}(1-v_{n-1})^j. 
\label{eq1}
\end{equation}
Nodes that have degree $i$ after pruning $n$ have degree $j$, which is
never less than $max\{i,k\}$, and $j-i$ neighbors are removed after
pruning $n$,
\begin{equation}
p_{n,i}=\sum_{j=max\{i, k\}}^{\infty}p_{n-1,j}{j \choose
  i}v_{n-1}^i(1-v_{n-1})^{j-i}.  
\label{eq2}
\end{equation}
And the relationship between the degree distribution and the excess degree distribution after pruning $n$ is:
\begin{equation}
q_{n,i}=\frac{(i+1)p_{n,i+1}}{\sum_{i=0}^{\infty}ip_{n,i}}.
\end{equation}

Although the theoretical framework for the four degree distribution
evolution equations was first proposed by Baxter et al.~\cite{baxter2015critical}, 
they noted that they were ``difficult to study analytically'' and solved them
``numerically for Erd\H{o}s-R\'enyi networks (Poisson degree
distributions) using the initial mean degree $\langle q\rangle$ as a
control parameter.'' To obtain the result, one must use the degree
distribution of the subgraph after the last pruning as an input, which
is an infinite-dimensional vector when the network is large.%
In what follows, by introducing an auxiliary series $y_n$ we can simplify the complex
infinite-dimensional simultaneous recurrence equations and make it an
equivalent univariable iteration process. Then the relevant quantities,
e.g., the size of the remaining subgraph $S_n$ in step $n$, can be obtained
and expressed as a simple function of $y_n$. 

First, we obtain the recurrence relation of $G_{n,0}(z)$ from \eqref{eq1}, and
\eqref{eq2} (see SI 1B for detail),
\begin{eqnarray}
G_{n,0}(z)=&&G_{n-1,0}(1-v_{n-1}+zv_{n-1})\nonumber\\
			&&+\sum_{j=0}^{k-1}p_{n-1,j}(1-(1-v_{n-1}+zv_{n-1})^j).
\end{eqnarray}
To acquire the general form of $G_{n,0}$, we introduce an important auxiliary series,
\begin{align}
y_0&=1,\\
y_{n}&=1-\sum_{j=0}^{k-2}\frac{y_{n-1}^j}{j!}G_{1}^{(j)}(1-y_{n-1})
\equiv f(y_{n-1}).
\end{align}
Here $G^{(j)}(z)$ is derivative $j$ of $G(z)$, i.e.,
$G^{(j)}(z)=d^jG(z)/dz^j$. Then by induction we have the generating function(see SI 1C for detail)
\begin{align}
&G_{n,0}(z)=G_0(1-y_n+y_nz)\nonumber\\
&+\sum_{j=0}^{k-1}\frac{G_0^{(j)}(1-y_{n-1})}{j!}(y_{n-1}^j-(y_{n-1}-y_n+y_nz)^j).
\label{generating}
\end{align}
The remaining subgraph after pruning $n$ have degrees no less than $k$
in the $\mathcal{N}_{n-1}$ network. Thus the size of the remaining
subgraph $S_n$ after pruning $n$ is
\begin{equation}
	S_n=\sum_{j=k}^{\infty}p_{n-1,j}
	=1-\sum_{j=0}^{k-1}\frac{G_0^{(j)}(1-y_{n-1})}{j!}y_{n-1}^j
	\equiv g(y_{n-1}).
	\end{equation}
Note that $S=\lim_{n \rightarrow \infty}S_n=g(y)$ is the size of the
final $k$-core, and that $y$ is the largest root of $y=f(y)$ in $[0,1)$. 
This result is consistent with previous research \cite{fernholz2004cores}.

\begin{figure*}[htb]
\centering
\includegraphics[scale=0.4]{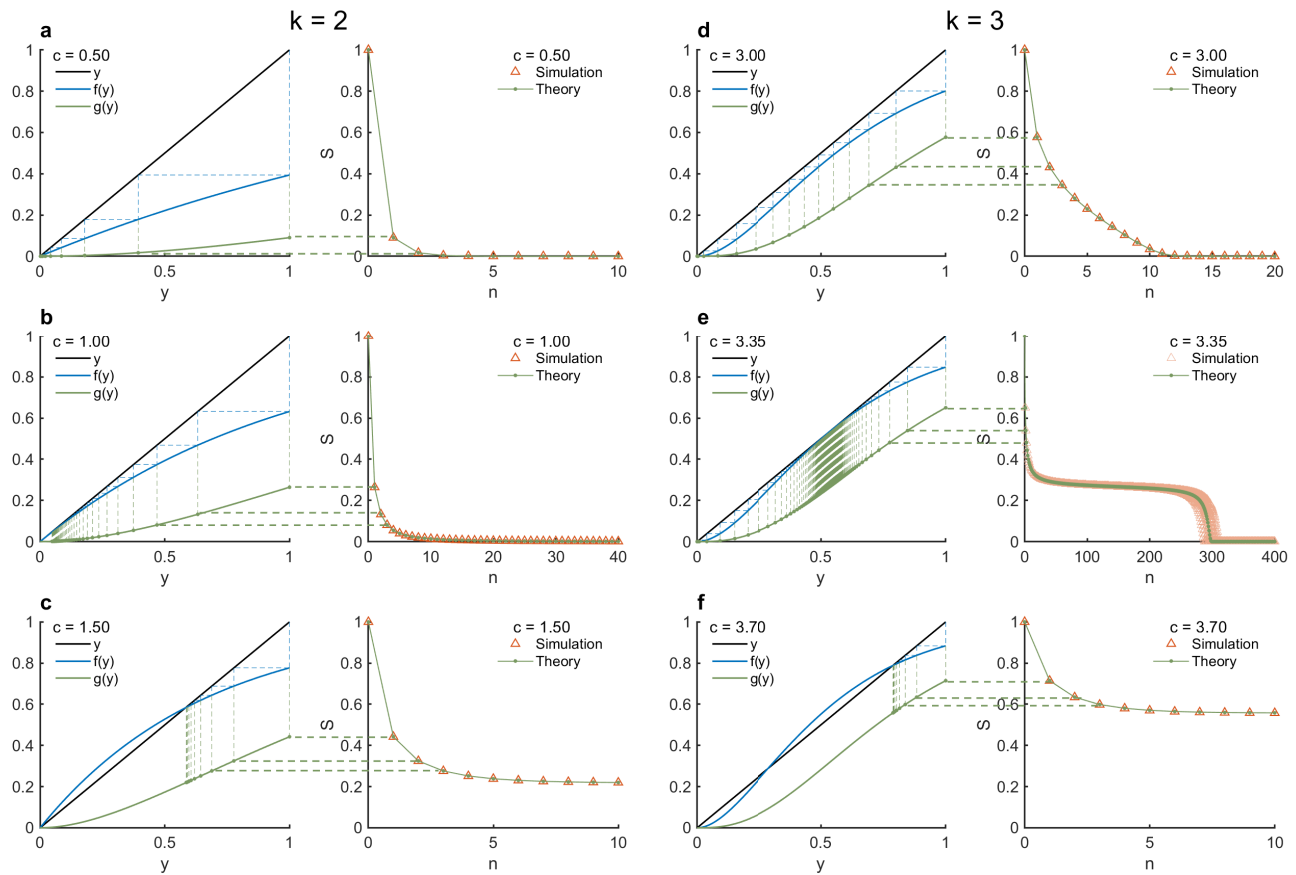}
\caption{
\textbf{An illustration for the size of remaining subgraph $S_n$ in $k$-core pruning process performed on ER-networks.} All panels consist of two panels, left and right. The left panels show $g(y)$ and the iteration of $y_n=f(y_{n-1})$ during the pruning process and the right panels show $S_n$ from both the theoretical result and numerical simulation result. The green dashed lines in between are indications of the corresponding relationship of $S_n=g(y_{n-1})$ according to our analytical result. \textbf{a-c)}, the result of $2$-core pruning process below, at, and above the critical point $c^*=1$, respectively. \textbf{d-f)}, the result of $3$-core pruning process below, near, and above the critical point $c^*=3.3509$, respectively. For \textbf{a,b,c,d,f}, the numerical results are obtained from simulations on $10^6$ nodes. Since the result is very sensitive to the random perturbation of the original network near the discontinuous phase transition point, the numerical result in \textbf{e} is averaged over 10 simulations on $5\times10^8$ nodes. 
}
\label{size}
\end{figure*}

\begin{figure*}[htb]
\centering
\includegraphics[scale=0.35]{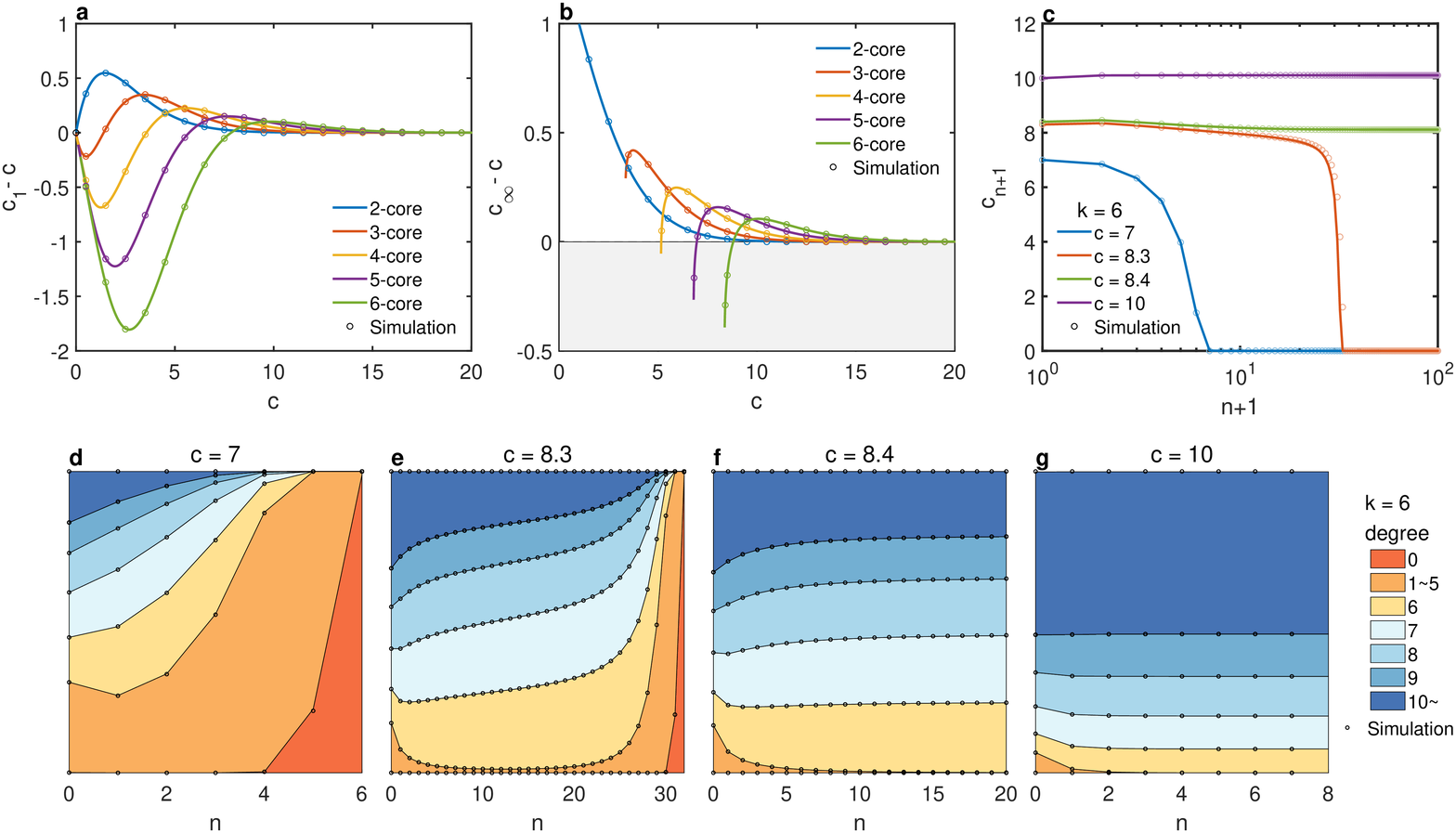}
\caption{
\textbf{The structural properties of remaining subgraph of $k$-core decomposition on ER-network.} The solid lines indicate the theoretical result and the circles represent the simulation result. \textbf{a)}, the average degree change after the first step of pruning $c_1-c$ versus the initial average degree $c$. \textbf{b)}, the difference between initial average degree and the average degree of $k$-core, denoted by $c_\infty-c$, versus the initial average degree $c$.  \textbf{c)}, The average degree evolution in each step of the $6$-core pruning process. Note that the horizontal axis is in logarithmic scale, and we shift $n^{th}$ step to $(n+1)^{th}$ step in order to show the data of initial condition ($n=0$). \textbf{d-g)}, the stacked area chart of the nodes with different degrees, evolves with pruning steps. Here simulations of $6$-core decomposition are performed on $4$ random ER-networks with different initial average degrees, the results are shown in \textbf{d-g}, respectively. The networks have $10^6$ nodes except in \textbf{e)}, the network has $10^8$ nodes. 
}
\label{structure}
\end{figure*}

From the exact expression of $S_n$ and $y_n$, we know that the
computation of $y_n$ is equivalent to a fixed-point iteration and that
$S_n$ can then be solved because it is a function of $y_{n-1}$. For
2-core decomposition on an ER-network, we obtain $f(y) = 1 - e^{-cy}$,
$g(y) = 1 - e^{-cy}(1+cy)$. Thus each step of the pruning process can be
represented by a corresponding iterative step of
$y=f(y)$. Fig.~\ref{size} shows the process using a simple
visualization method.

In 3-core decomposition, it is easy to acquire $f(y) = 1 -
e^{-cy}(1+cy)$ and $g(y) = 1 - e^{-cy}(1+cy+(cy)^2/2)$. Unlike the
result from 2-core decomposition, there is a discontinuous phase
transition at the critical point $c^* = 3.3509$ (see
Fig.~\ref{size}\textbf{d-f}). The pruning process exhibits interesting
behavior when $c$ approaches the critical point from the left (see
Fig.~\ref{size}\textbf{e}). In the first few pruning steps, $S_n$
rapidly decreases. The pruning then reaches a bottleneck, then becomes
transient process, then experiences an avalanche of node removal. This
phenomenon was observed in previous research
\cite{baxter2015critical}. Our analytical result explains this
interesting discontinuous phase transition.  When $c\ll c^*$ the
iteration quickly converges to a stable fixed point at $y=0$ (see
Fig.~\ref{size}\textbf{d}), and thus no $k$-core remains. When $c>c^*$
the iteration stops at the largest root of $y^*=f(y^*)$. Between those
two scenarios, when $c$ approaches $c^*$ from the left (see
Fig.~\ref{size}\textbf{e}), the curve of $f(y)$ and the diagonal line
together form a long narrow tube through which the iteration process
passes slowly but does not stop. After passing through the narrow tube
it stops at a stable fixed point at $y=0$, which is in accord with the
critical phenomena described above.

From the generating function (Eq.~\ref{generating}), we can easily obtain the degree distribution of the remaining subgraph $\mathcal{N}_n$ in $n^{th}$ step, which depicts the detailed topological structure of the intermediate state of the network. Here we show the average degree of $\mathcal{N}_n$, $c_n=G'_{n,0}(1)/S_n=cy_n^2/S_n$ (See SI 2A) as an example.

We examine $c_1$ and $c_\infty$, which are two representative
examples. Fig.~\ref{structure}\textbf{a,b} shows our analytical and
simulation results. Inset \textbf{a} shows how when the initial average
degree varies in $k$-core decomposition the average degree changes after
the first pruning step. Note that the average degree increases or
decreases depending on the initial average degree. Inset \textbf{b}
shows the average degree change corresponding to the final $k$-core. One
counter-intuitive phenomenon is that the average $k$-core degree can be
smaller than the original network (shaded area in \textbf{b}). A special
case of $7$-core decomposition on an ER-network when $c=10$ was also
recently described by Yoon et al.~\cite{yoon2018structural}.
Fig.~\ref{structure}\textbf{c} shows the $6$-core pruning process. When
$c=8.4$, a value slightly above the critical point, the average degree
of the $6$-core is lower than that in the original
network. Fig.~\ref{structure}\textbf{d-g} also shows the stacked area chart
illustrating the detailed evolution of different degree
compositions. Note that when $c$ is slightly above the critical point
$c^*$, most of the $6$-core is composed of nodes with degrees of $6$,
$7$, and $8$. This explains why under certain conditions the average
degree of the $k$-core can be smaller than the original average
degree. We examined above $k$-core decomposition on ER-networks. We
also use our method to examine the $k$-core decomposition of scale-free
networks, and the simulations further validate our theoretical results
(see SI 4).

To summarize, we have studied the $k$-core pruning process and obtained an analytical solution that describes the whole pruning process. By introducing auxiliary series $y_n$, we simplify the existing theoretical framework \cite{baxter2015critical} to a simple univariable iteration and thus are able to obtain the analytical solution. We also obtain the results of complete evolution process including the size and structure of the remaining subgraph. Numerical simulations confirm that our analytical results are solid. Our major contribution here is that we develop an new method of greatly simplifying and reforming the way we understand $k$-core decomposition in any large uncorrelated network. We describe the precise critical behavior of the high dimensional interacting system by mapping it to a simple univariable iteration process. This simplification can serve as a powerful for further research in a variety of related fields.

\section*{Acknowledgment}
We thank Prof. Matus Medo and Dr. Chi Zhang for helpful discussions. We give special thanks to Dr. G. J. Baxter, Prof. S. N. Dorogovtsev, Dr. K.-E. Lee, Prof. J. F. F. Mendes, and Prof. A. V. Goltsev for valuable comments.

\bibliographystyle{apsrev4-1}
\bibliography{bibliography}

\end{document}